\documentclass[%
reprint,
superscriptaddress,
 amsmath,amssymb,
 aps,
floatfix,
]{revtex4-2}

\usepackage{graphicx}
\usepackage{dcolumn}
\usepackage{bm}
\input{symbols}
\usepackage{color,soul} 
\usepackage{multirow}

\begin{document}


%
%
\newcommand{\pvalRateOnlyBetaZeroDataRunOneNineD}{$0.059$}
\newcommand{\sigmaRateOnlyBetaZeroDataRunOneNineD}{$1.89$}
\newcommand{\pvalRateOnlyBetaOneDataRunOneNineD}{$0.321$}
\newcommand{\sigmaRateOnlyBetaOneDataRunOneNineD}{$0.99$}
\newcommand{\pvalRateShapeBetaZeroDataRunOneNineD}{$0.016$}
\newcommand{\sigmaRateShapeBetaZeroDataRunOneNineD}{$2.40$}
\newcommand{\pvalRateShapeBetaOneDataRunOneNineD}{$0.300$}
\newcommand{\sigmaRateShapeBetaOneDataRunOneNineD}{$1.04$}
\newcommand{\DeltachisqDataRunOneNineD}{3.811}
\newcommand{\BayesFactorDataRunOneNineD}{0.149}
\newcommand{\NskObservedNueBarRunOneNineD}{15}

%
%
\newcommand{\pvalRateOnlyBetaZeroAsimovARunOneNineD}{$0.019$}
\newcommand{\sigmaRateOnlyBetaZeroAsimovARunOneNineD}{$2.36$}
\newcommand{\pvalRateOnlyBetaOneAsimovARunOneNineD}{$0.379$}
\newcommand{\sigmaRateOnlyBetaOneAsimovARunOneNineD}{$0.88$}
\newcommand{\pvalRateShapeBetaZeroAsimovARunOneNineD}{$0.006$}
\newcommand{\sigmaRateShapeBetaZeroAsimovARunOneNineD}{$2.76$}
\newcommand{\pvalRateShapeBetaOneAsimovARunOneNineD}{$0.409$}
\newcommand{\sigmaRateShapeBetaOneAsimovARunOneNineD}{$0.83$}
\newcommand{\DeltachisqAsimovRunOneNineD}{6.3}
\newcommand{\NskAsimovNueBarRunOneNineD}{16.8}

\preprint{APS/123-QED}

\title{Search for Electron Antineutrino Appearance in a Long-baseline Muon Antineutrino Beam}


\newcommand{\INSTHD}{\affiliation{University Autonoma Madrid, Department of Theoretical Physics, 28049 Madrid, Spain}}
\newcommand{\INSTEE}{\affiliation{University of Bern, Albert Einstein Center for Fundamental Physics, Laboratory for High Energy Physics (LHEP), Bern, Switzerland}}
\newcommand{\INSTFE}{\affiliation{Boston University, Department of Physics, Boston, Massachusetts, U.S.A.}}
\newcommand{\INSTD}{\affiliation{University of British Columbia, Department of Physics and Astronomy, Vancouver, British Columbia, Canada}}
\newcommand{\INSTGA}{\affiliation{University of California, Irvine, Department of Physics and Astronomy, Irvine, California, U.S.A.}}
\newcommand{\INSTI}{\affiliation{IRFU, CEA Saclay, Gif-sur-Yvette, France}}
\newcommand{\INSTGB}{\affiliation{University of Colorado at Boulder, Department of Physics, Boulder, Colorado, U.S.A.}}
\newcommand{\INSTFG}{\affiliation{Colorado State University, Department of Physics, Fort Collins, Colorado, U.S.A.}}
\newcommand{\INSTFH}{\affiliation{Duke University, Department of Physics, Durham, North Carolina, U.S.A.}}
\newcommand{\INSTBA}{\affiliation{Ecole Polytechnique, IN2P3-CNRS, Laboratoire Leprince-Ringuet, Palaiseau, France }}
\newcommand{\INSTEF}{\affiliation{ETH Zurich, Institute for Particle Physics and Astrophysics, Zurich, Switzerland}}
\newcommand{\INSTIE}{\affiliation{CERN European Organization for Nuclear Research, CH-1211 Genève 23, Switzerland}}
\newcommand{\INSTEG}{\affiliation{University of Geneva, Section de Physique, DPNC, Geneva, Switzerland}}
\newcommand{\INSTHJ}{\affiliation{University of Glasgow, School of Physics and Astronomy, Glasgow, United Kingdom}}
\newcommand{\INSTDG}{\affiliation{H. Niewodniczanski Institute of Nuclear Physics PAN, Cracow, Poland}}
\newcommand{\INSTCB}{\affiliation{High Energy Accelerator Research Organization (KEK), Tsukuba, Ibaraki, Japan}}
\newcommand{\INSTIB}{\affiliation{University of Houston, Department of Physics, Houston, Texas, U.S.A.}}
\newcommand{\INSTED}{\affiliation{Institut de Fisica d'Altes Energies (IFAE), The Barcelona Institute of Science and Technology, Campus UAB, Bellaterra (Barcelona) Spain}}
\newcommand{\INSTEC}{\affiliation{IFIC (CSIC \& University of Valencia), Valencia, Spain}}
\newcommand{\INSTHH}{\affiliation{Institute For Interdisciplinary Research in Science and Education (IFIRSE), ICISE, Quy Nhon, Vietnam}}
\newcommand{\INSTEI}{\affiliation{Imperial College London, Department of Physics, London, United Kingdom}}
\newcommand{\INSTGF}{\affiliation{INFN Sezione di Bari and Universit\`a e Politecnico di Bari, Dipartimento Interuniversitario di Fisica, Bari, Italy}}
\newcommand{\INSTBE}{\affiliation{INFN Sezione di Napoli and Universit\`a di Napoli, Dipartimento di Fisica, Napoli, Italy}}
\newcommand{\INSTBF}{\affiliation{INFN Sezione di Padova and Universit\`a di Padova, Dipartimento di Fisica, Padova, Italy}}
\newcommand{\INSTBD}{\affiliation{INFN Sezione di Roma and Universit\`a di Roma ``La Sapienza'', Roma, Italy}}
\newcommand{\INSTEB}{\affiliation{Institute for Nuclear Research of the Russian Academy of Sciences, Moscow, Russia}}
\newcommand{\INSTHI}{\affiliation{International Centre of Physics, Institute of Physics (IOP), Vietnam Academy of Science and Technology (VAST), 10 Dao Tan, Ba Dinh, Hanoi, Vietnam}}
\newcommand{\INSTHA}{\affiliation{Kavli Institute for the Physics and Mathematics of the Universe (WPI), The University of Tokyo Institutes for Advanced Study, University of Tokyo, Kashiwa, Chiba, Japan}}
\newcommand{\INSTID}{\affiliation{Keio University, Department of Physics, Kanagawa, Japan}}
\newcommand{\INSTIF}{\affiliation{King's College London, Department of Physics, Strand, London WC2R 2LS, United Kingdom}}
\newcommand{\INSTCC}{\affiliation{Kobe University, Kobe, Japan}}
\newcommand{\INSTCD}{\affiliation{Kyoto University, Department of Physics, Kyoto, Japan}}
\newcommand{\INSTEJ}{\affiliation{Lancaster University, Physics Department, Lancaster, United Kingdom}}
\newcommand{\INSTFC}{\affiliation{University of Liverpool, Department of Physics, Liverpool, United Kingdom}}
\newcommand{\INSTFI}{\affiliation{Louisiana State University, Department of Physics and Astronomy, Baton Rouge, Louisiana, U.S.A.}}
\newcommand{\INSTHB}{\affiliation{Michigan State University, Department of Physics and Astronomy,  East Lansing, Michigan, U.S.A.}}
\newcommand{\INSTCE}{\affiliation{Miyagi University of Education, Department of Physics, Sendai, Japan}}
\newcommand{\INSTDF}{\affiliation{National Centre for Nuclear Research, Warsaw, Poland}}
\newcommand{\INSTFJ}{\affiliation{State University of New York at Stony Brook, Department of Physics and Astronomy, Stony Brook, New York, U.S.A.}}
\newcommand{\INSTGJ}{\affiliation{Okayama University, Department of Physics, Okayama, Japan}}
\newcommand{\INSTCF}{\affiliation{Osaka City University, Department of Physics, Osaka, Japan}}
\newcommand{\INSTGG}{\affiliation{Oxford University, Department of Physics, Oxford, United Kingdom}}
\newcommand{\INSTGC}{\affiliation{University of Pittsburgh, Department of Physics and Astronomy, Pittsburgh, Pennsylvania, U.S.A.}}
\newcommand{\INSTFA}{\affiliation{Queen Mary University of London, School of Physics and Astronomy, London, United Kingdom}}
\newcommand{\INSTE}{\affiliation{University of Regina, Department of Physics, Regina, Saskatchewan, Canada}}
\newcommand{\INSTGD}{\affiliation{University of Rochester, Department of Physics and Astronomy, Rochester, New York, U.S.A.}}
\newcommand{\INSTHC}{\affiliation{Royal Holloway University of London, Department of Physics, Egham, Surrey, United Kingdom}}
\newcommand{\INSTBC}{\affiliation{RWTH Aachen University, III. Physikalisches Institut, Aachen, Germany}}
\newcommand{\INSTFB}{\affiliation{University of Sheffield, Department of Physics and Astronomy, Sheffield, United Kingdom}}
\newcommand{\INSTDI}{\affiliation{University of Silesia, Institute of Physics, Katowice, Poland}}
\newcommand{\INSTIA}{\affiliation{SLAC National Accelerator Laboratory, Stanford University, Menlo Park, California, USA}}
\newcommand{\INSTBB}{\affiliation{Sorbonne Universit\'e, Universit\'e Paris Diderot, CNRS/IN2P3, Laboratoire de Physique Nucl\'eaire et de Hautes Energies (LPNHE), Paris, France}}
\newcommand{\INSTEH}{\affiliation{STFC, Rutherford Appleton Laboratory, Harwell Oxford,  and  Daresbury Laboratory, Warrington, United Kingdom}}
\newcommand{\INSTCH}{\affiliation{University of Tokyo, Department of Physics, Tokyo, Japan}}
\newcommand{\INSTBJ}{\affiliation{University of Tokyo, Institute for Cosmic Ray Research, Kamioka Observatory, Kamioka, Japan}}
\newcommand{\INSTCG}{\affiliation{University of Tokyo, Institute for Cosmic Ray Research, Research Center for Cosmic Neutrinos, Kashiwa, Japan}}
\newcommand{\INSTHF}{\affiliation{Tokyo Institute of Technology, Department of Physics, Tokyo, Japan}}
\newcommand{\INSTGI}{\affiliation{Tokyo Metropolitan University, Department of Physics, Tokyo, Japan}}
\newcommand{\INSTHG}{\affiliation{Tokyo University of Science, Faculty of Science and Technology, Department of Physics, Noda, Chiba, Japan}}
\newcommand{\INSTF}{\affiliation{University of Toronto, Department of Physics, Toronto, Ontario, Canada}}
\newcommand{\INSTB}{\affiliation{TRIUMF, Vancouver, British Columbia, Canada}}
\newcommand{\INSTG}{\affiliation{University of Victoria, Department of Physics and Astronomy, Victoria, British Columbia, Canada}}
\newcommand{\INSTDJ}{\affiliation{University of Warsaw, Faculty of Physics, Warsaw, Poland}}
\newcommand{\INSTDH}{\affiliation{Warsaw University of Technology, Institute of Radioelectronics and Multimedia Technology, Warsaw, Poland}}
\newcommand{\INSTFD}{\affiliation{University of Warwick, Department of Physics, Coventry, United Kingdom}}
\newcommand{\INSTGH}{\affiliation{University of Winnipeg, Department of Physics, Winnipeg, Manitoba, Canada}}
\newcommand{\INSTEA}{\affiliation{Wroclaw University, Faculty of Physics and Astronomy, Wroclaw, Poland}}
\newcommand{\INSTHE}{\affiliation{Yokohama National University, Faculty of Engineering, Yokohama, Japan}}
\newcommand{\INSTH}{\affiliation{York University, Department of Physics and Astronomy, Toronto, Ontario, Canada}}

\INSTHD
\INSTEE
\INSTFE
\INSTD
\INSTGA
\INSTI
\INSTGB
\INSTFG
\INSTFH
\INSTBA
\INSTEF
\INSTIE
\INSTEG
\INSTHJ
\INSTDG
\INSTCB
\INSTIB
\INSTED
\INSTEC
\INSTHH
\INSTEI
\INSTGF
\INSTBE
\INSTBF
\INSTBD
\INSTEB
\INSTHI
\INSTHA
\INSTID
\INSTIF
\INSTCC
\INSTCD
\INSTEJ
\INSTFC
\INSTFI
\INSTHB
\INSTCE
\INSTDF
\INSTFJ
\INSTGJ
\INSTCF
\INSTGG
\INSTGC
\INSTFA
\INSTE
\INSTGD
\INSTHC
\INSTBC
\INSTFB
\INSTDI
\INSTIA
\INSTBB
\INSTEH
\INSTCH
\INSTBJ
\INSTCG
\INSTHF
\INSTGI
\INSTHG
\INSTF
\INSTB
\INSTG
\INSTDJ
\INSTDH
\INSTFD
\INSTGH
\INSTEA
\INSTHE
\INSTH

\author{K.\,Abe}\INSTBJ
\author{R.\,Akutsu}\INSTCG
\author{A.\,Ali}\INSTCD
\author{C.\,Alt}\INSTEF
\author{C.\,Andreopoulos}\INSTEH\INSTFC
\author{L.\,Anthony}\INSTFC
\author{M.\,Antonova}\INSTEC
\author{S.\,Aoki}\INSTCC
\author{A.\,Ariga}\INSTEE
\author{Y.\,Asada}\INSTHE
% ...

\collaboration{The T2K Collaboration}\noaffiliation

\date{\today}

\begin{abstract}
Electron antineutrino appearance is measured by the T2K experiment in an accelerator-produced antineutrino beam, using additional neutrino beam operation to constrain parameters of the PMNS mixing matrix. T2K observes \NskObservedNueBarRunOneNineD\ candidate electron antineutrino events with a background expectation of 9.3 events. Including information from the kinematic distribution of observed events, the hypothesis of no electron antineutrino appearance is disfavored with a significance of \sigmaRateShapeBetaZeroDataRunOneNineD$\sigma$ and no discrepancy between data and PMNS predictions is found. A complementary analysis that introduces an additional free parameter which allows non-PMNS values of electron neutrino and antineutrino appearance also finds no discrepancy between data and PMNS predictions. 
\end{abstract}

\maketitle

{\it Introduction}\textemdash The observation of neutrino oscillations has established that each neutrino flavor state ($e$, $\mu$, $\tau$) is a superposition of at least three mass eigenstates ($m_1$, $m_2$, $m_3$)~\cite{PhysRevLett.81.1562,PhysRevLett.87.071301,PhysRevLett.107.041801,PhysRevLett.108.171803}. The phenomenon of oscillation is modeled by a three-generation flavor-mass mixing matrix, called the Pontecorvo-Maki-Nakagawa-Sakata (PMNS) matrix~\cite{maki1962remarks,pontecorvo1968neutrino}. With the discovery of non-zero $\theta_{13}$ and the explicit observation of \num to \nue appearance oscillation~\cite{abe2014observation}, it is now crucial to test the PMNS framework and establish if it is sufficient to explain all neutrino and antineutrino oscillation observations. One such test is to search for the $CP$-reversed appearance oscillation of \numb to \nueb. A search for this process in the Tokai-to-Kamioka (T2K) experiment was reported in reference~\cite{PhysRevD.96.092006}, and recent results from the NOvA experiment show a significance of 4.4$\sigma$~\cite{acero2019first}.  In this Letter, we report a search for electron antineutrino appearance at the T2K experiment with an improved event selecton and a dataset more than a factor of two larger than previous T2K results.


{\it The T2K Experiment}\textemdash  The T2K experiment~\cite{abe2011t2k} begins with a 30~GeV proton beam from the J-PARC main ring striking a graphite target, producing pions and kaons. These charged hadrons are focused by a system of three magnetic horns to decay in a 96~m decay volume. Positively charged hadrons are focused to produce a beam of predominantly neutrinos (``neutrino mode"); negatively charged hadrons are focused for a beam of predominantly antineutrinos (``antineutrino mode"). 

An unmagnetized on-axis near detector (INGRID) and a magnetized off-axis (2.5$^{\circ}$) near detector (ND280) sample the unoscillated neutrino beam 280 m downstream from the target station and monitor the beam direction, composition, and intensity and constrain neutrino interaction properties. The unmagnetized Super-Kamiokande (SK) 50 kt water-Cherenkov detector is the T2K far detector, and samples the oscillated neutrino beam 2.5$^{\circ}$ off axis and 295~km from the production point.

The analysis presented here uses data collected from January 2010 to June 2018. The data set has an exposure at SK of $1.63\times10^{21}$ protons on target (POT) in antineutrino mode, with an additional data set of $1.49\times10^{21}$ POT in neutrino mode used to constrain PMNS oscillation parameters acting as systematic uncertainties in the analysis. The ND280 detector uses an exposure of $0.58\times10^{21}$ POT in neutrino mode and $0.39\times10^{21}$ POT in antineutrino mode.

{\it Analysis Strategy}\textemdash The significance of $\nueb$ appearance is evaluated by introducing the parameter $\beta$, which multiplies the PMNS oscillation probability $P\left(\numb \rightarrow \nueb\right)$:
\begin{equation}
P\left(\numb \rightarrow \nueb\right) = \beta \times P_{\mbox{PMNS}}\left(\numb \rightarrow \nueb\right)
\label{eq:nueb}
\end{equation}

The analysis is performed allowing both $\beta=0$ and $\beta=1$ to be the null hypothesis, where both hypotheses fully account for uncertainties in the values of the oscillation and systematic parameters. Two analyses are performed on each hypothesis to obtain corresponding p-values: one uses only the number of events (‘rate-only’); while the other also uses information from the kinematic variables of events (‘rate+shape’).

The total number of candidate \nueb events in the antineutrino beam mode is used as the test statistic to calculate the rate-only $p$-value. The test statistic 
\begin{equation}
\Delta \chi^2 = \chi^2 \left( \beta = 0 \right) - \chi^2 \left( \beta = 1 \right) 
\label{eq:dchisq01}
\end{equation}
is used to calculate the rate+shape $p$-value, where the $\chi^2$ values are calculated by marginalizing over all systematic and oscillation parameters, including the mass hierarchy. In both analyses, other data samples---\num-like and \nue-like in neutrino beam mode and \numb-like in antineutrino beam mode---are used to constrain other PMNS oscillation parameters, as in other T2K analyses~\cite{PhysRevLett.121.171802}. 

A complementary analysis allows $\beta$ to be a continuous free parameter with limits between 0 and infinity. In this analysis only, in addition to $\beta$ multiplying $P_{\mbox{PMNS}}\left(\numb \rightarrow \nueb\right)$ as in Eq.~\ref{eq:nueb}, the probability $P_{\mbox{PMNS}}\left(\num \rightarrow \nue\right)$ is multiplied by a factor $1/\beta$. The analysis results are independent of whether $1/\beta$ multiplies the neutrino oscillation probability; it is used for symmetry. The extra degree of freedom allows the fit to explore areas away from the PMNS constraint to more accurately reflect the information given by the data. Credible interval contours in the \pnue and \pnueb parameter space, the main result of the analysis, are then compared against T2K data fit with $\beta$  fixed to 1  to test the compatibility between the T2K data and the PMNS model constraining the standard fit.

{\it Neutrino Beam Flux}\textemdash The primary signal data sets were taken in antineutrino mode. The flux was predicted by a Monte Carlo (MC) simulation incorporating the FLUKA2011 interaction model \cite{Ferrari:2005zk} tuned to the results of recent external hadron production experiments including the NA61/SHINE experiment at CERN \cite{Abgrall:2011ae,PhysRevC.85.035210,Abgrall:2016jif}. The INGRID detector is used to monitor the beam axis direction and total flux stability.

The resultant flux model \cite{Abe:2012av,Posiadala-Zezula:2017ivt,zambelli2017towards} estimates unoscillated neutrino and antineutrino fluxes at all detectors as well as their uncertainties and correlations. The flux at ND280 and SK peaks at 600~MeV, where 96.2\% of the beam is composed of $\bar\nu_\mu$ and 0.46\% $\bar\nu_e$. The remainder of the beam is almost entirely $\nu_\mu$. This wrong sign contamination is greater in antineutrino mode than neutrino mode.


{\it Neutrino Interaction Model}\textemdash The NEUT (v5.3.3) neutrino interaction generator~\cite{hayato2009neutrino} is used to generate simulated neutrino events. The model used is described in references~\cite{PhysRevD.96.092006} and~\cite{PhysRevLett.121.171802}. The most relevant contributions for this analysis are highlighted here. 

The dominant neutrino-nucleus interaction topology near 600 MeV, charged current quasielastic (CCQE)-like, is defined as an interaction with one charged lepton and zero pions in the final state. The nucleus is modeled with a relativistic Fermi gas (RFG) modified by a random phase approximation (RPA) to account for long-range correlations~\cite{PhysRevC.70.055503}. A multinucleon component is included with the Nieves 2$p$-2$h$ model~\cite{PhysRevD.88.113007,PhysRevC.83.045501}, which contains both meson exchange current ($\Delta$-like) and correlated nucleon pair (non-$\Delta$-like) contributions. Parameters representing systematic uncertainties for the CCQE-like mode include the nucleon axial mass, $M_A^{QE}$; the Fermi momentum for $^{12}$C and $^{16}$O; the 2$p$-2$h$ normalization term for $\nu$ and \nub separately; four parameters controlling the RPA shape as a function of $Q^2$; and the relative contributions of the $\Delta$-like and non-$\Delta$-like contributions to 2$p$-2$h$ in $^{12}$C and $^{16}$O. The RPA parameters have Gaussian priors to cover the theoretical shape uncertainty given in~\cite{VALVERDE2006325,gran2017model}, and the 2$p$-2$h$ shape contribution has a 30\% correlation between $^{12}$C and $^{16}$O; all other priors are uniform. Other neutrino-nucleus processes are subdominant, and their rates are constrained via appropriate uncertainties. 


Differences between muon- and electron-neutrino interactions are largest at low energies and occur because of final-state lepton mass and radiative corrections. A 2\% uncorrelated uncertainty is added for each of the electron neutrino and antineutrino cross sections relative to those of muons and another 2\% uncertainty anticorrelated between the two ratios~\cite{PhysRevD.86.053003}. 

Some systematic uncertainties are not easily included by varying model parameters. These are the subjects of ``simulated data'' studies, where simulated data generated from a variant model are analyzed under the assumptions of the default model. The model variations that produce the largest changes in the \nueb far detector spectra are an alternate single resonant pion model~\cite{PhysRevD.97.013002}, and ad-hoc models driven by observed discrepancies in the near detector kinematic spectra, where the discrepancy is modeled as having either 1$p$-1$h$, 2$p$-2$h$-$\Delta$-like, and 2$p$-2$h$-non-$\Delta$-like kinematics. None of the variant models studied showed differences in the sensitivity values at greater than the 0.1$\sigma$ level. 

{\it Near Detector Data Constraints}\textemdash The ND280 detector is used to fit unoscillated samples of charged current (CC) muon neutrino interaction events to constrain flux and cross section systematic uncertainties for the signal and background models of SK events. The samples---unchanged from reference~\cite{PhysRevLett.121.171802}---are selected from events that begin in one of two fine-grained detectors (FGDs) and produce tracks that enter the time-projection chambers (TPCs), which are interleaved with the FGDs. Both FGDs are composed of layers of bars of plastic scintillator, and the more downstream FGD additionally has panels of water interleaved between layers of scintillator. 

In neutrino beam mode, in each FGD, the CC events (defined as containing negatively charged muon-like track) are split into three subsamples: a CC0$\pi$ sample, with zero pions in the final state, enhanced in CCQE-like interactions; a CC1$\pi^{+}$ sample, with one $\pi^{+}$ in the final state, enhanced in resonant pion interactions; and a CC Other sample, containing all other CC events. In antineutrino beam mode, in each FGD, there are selected interactions with positively charged muons (\nub-like) and negatively charged muons ($\nu$-like). The latter constrains the wrong-sign contamination, which is higher in antineutrino beam mode. Each of these selections is divided into two topologies: containing a single track and containing multiple tracks.

All samples are fit simultaneously and are binned in lepton momentum, $p_{\mu}$, and lepton angle, $\cos\theta_{\mu}$ relative to the average beam neutrino direction. A binned likelihood fit to the data is performed assuming a Poisson-distributed number of events in each bin with an expectation computed from the flux, cross section, and ND280 detector models. The fit returns central values and correlated uncertainties for systematic uncertainty parameters that are constrained by the near detector, marginalizing over near detector flux and detector systematic parameters. Some uncertainties on neutral current and \nue events cannot be constrained by these ND280 samples and those parameters are passed to the appearance analysis with their original prior. 

The MC prediction before fitting underestimates the data by 10-15\%, consistent with previous T2K analyses. The agreement between the MC prediction after fitting and data is good, with a $p$-value of 0.473. The fit to the ND280 data reduces the flux and the ND280-constrained interaction model uncertainties on the predicted electron antineutrino sample event rate at the far detector from 14.6\% to 7.6\%. 


{\it \nueb SK selection}\textemdash  Unlike in the previous analysis, SK events are reconstructed and selected using the new reconstruction algorithm described in reference~\cite{Abe:2018wpn}. A $\bar\nu_e$ event candidate in SK must meet the following criteria: 1) it is within the beam time window as determined from a GPS time stamp, and its Cherenkov light is fully contained in the
SK inner detector, with minimal outer-detector activity; 2) the
reconstructed vertex is at least 80~cm from the inner-detector wall;
3) only one Cherenkov ring candidate is found in the reconstruction and
the ring is identified as electron-like; 4) the distance from the
vertex to the detector wall is greater than 170~cm along the track
direction; 5) the visible energy in the event is greater than 100~MeV;
6) there is no evidence of delayed activity consistent with a stopped
muon decay; 7) the reconstructed energy under a quasielastic
scattering hypothesis is less than 1250~MeV; 8) the ring is inconsistent with a $\pi^0$ decay hypothesis.

These reconstruction cuts have an efficiency of 71.5\% for $\bar\nu_e$
events that satisfy the fully-contained and fiducial requirements. The new event selection increases the yield of $\bar{\nu_e}$ signal by approximately 20\% compared to the previous analysis, primarily due to the new fiducial cuts, with no loss of purity. Assuming oscillation parameter values near the best fit of previous T2K analyses of $\sin^2 \theta_{23} = 0.528$, $\sin^2 \theta_{13} = 0.0212$, $\sin^2 \theta_{12} = 0.304$, $\Delta m^2_{32} = 2.509 \times 10^{-3} \mbox{ eV}^2 / c^4$, $\Delta m^2_{21} = 7.53 \times 10^{-5} \mbox{ eV}^2 / c^4$, $\delta_{CP} = -1.601$, normal hierarchy and $\beta = 1$, the total expected background is 9.3~events including 3.0 $\nu_e$
interactions resulting from oscillations of $\nu_\mu$ in the beam.
The remaining major sources of background are intrinsic $\nu_e$ and
$\bar\nu_e$ in the beam (4.2~events) and neutral-current interactions
(2.1~events).  With the oscillation parameters above, a signal yield
of 7.4~events is expected, for a total prediction of~\NskAsimovNueBarRunOneNineD~events.

\begin{figure}[ht]
	\includegraphics[width=\linewidth]{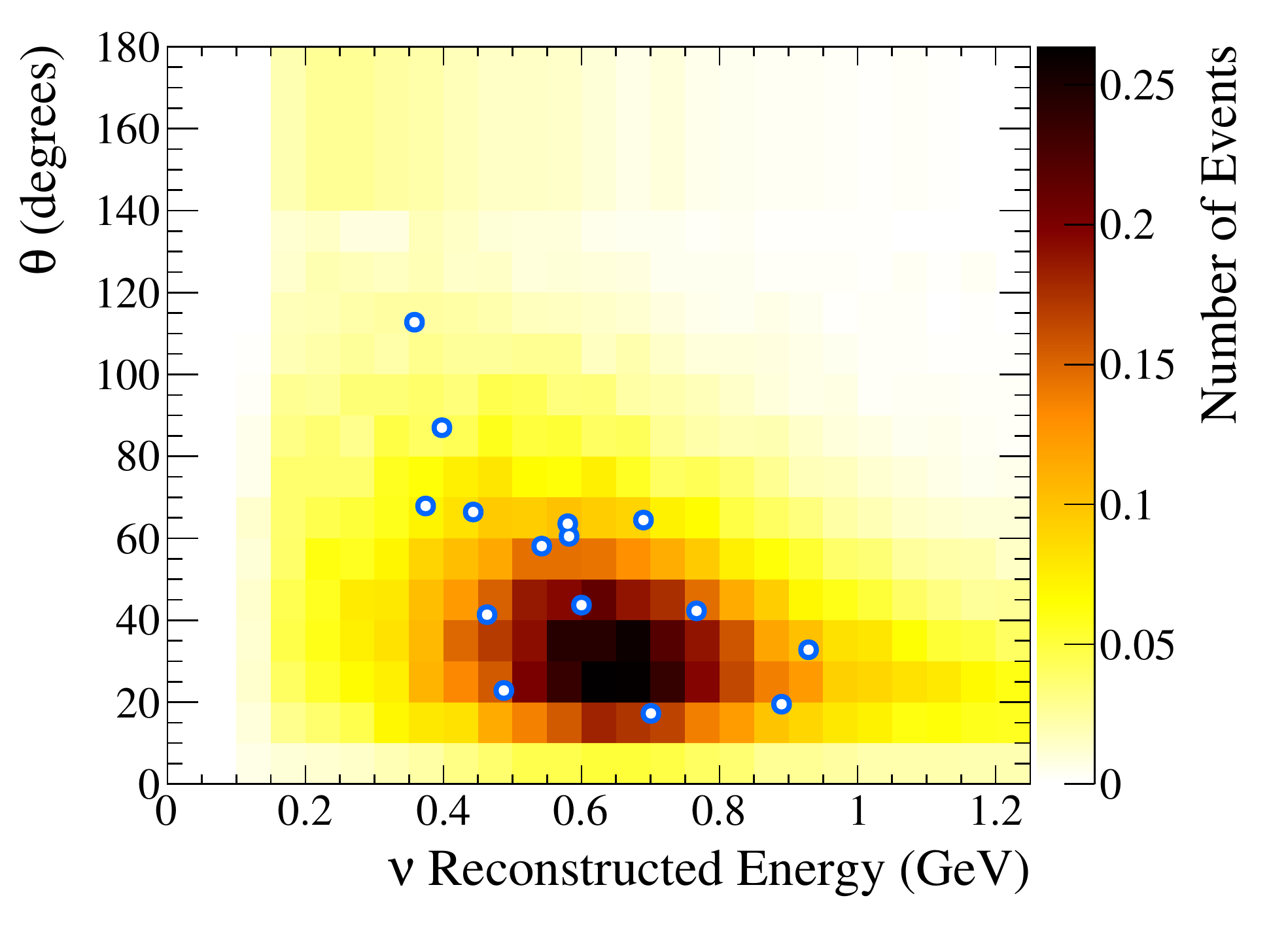}
	\hspace*{-0.5cm}
	\caption{Predicted $\nub$ mode single-ring e-like spectrum (coloured histogram) compared against T2K data (white/blue points). The distribution is a function of both the reconstructed neutrino energy and the reconstructed angle between the outgoing lepton and the neutrino direction.}
	\label{fig:spectrum_RHC1Re}
\end{figure}

Fig.~\ref{fig:spectrum_RHC1Re} shows the fifteen observed data events superimposed on a prediction generated using the above oscillation parameter values.

{\it \nueb Appearance}\textemdash The $\nueb$ appearance $p$-values are calculated by considering the rate-only and rate+shape test statistics of an ensemble of $2 \times 10^4$ pseudo-experiments. Each pseudo-experiment is generated by randomizing systematic parameters--including oscillation parameters--and applying statistical fluctuations. Four control samples, $\nu$ mode single-ring e-like and $\nue\ CC1\pi$-like (single-ring e-like accompanied by electron decay) and both $\nu$ and $\nub$ mode single-ring $\mu$-like, are used to constrain the distribution of oscillation parameters of the pseudo-experiments. The four control samples of many pseudo-experiments are compared to data, and rejection sampling is used to select 2$\times 10^4$ that are most probable, according to data. The systematic parameters are then marginalized over using a numeric integration technique (with 2$\times 10^5$ samples of the systematic parameter space) when calculating the rate+shape test statistic. Both the number of pseudo-experiments and the number of points used for the numerical integration were studied and selected to ensure $p$-value stability.

When producing the pseudo-experiments and marginalizing over systematic uncertainties, Gaussian prior probabilities on the following oscillation parameters are used: $\sin^{2} 2 \theta_{12}$ ($0.846 \pm 0.021$); $\Delta m^{2}_{21}$ $\left((7.53 \pm 0.18) \times 10^{-5}~\mbox{eV}^2 /\mbox{c}^4 \right)$; and  $\sin^{2} 2 \theta_{13}$ ($0.0830 \pm 0.0031$)
\cite{PDG2018}. The mass ordering is randomized with a probability of 0.5 for NO, 0.5 for IO. The other PMNS parameters are randomized using uniform prior probabilities with limits set based on previous experiments. Systematic parameters are randomized according to the constraints set by the near detector fit. 

When predicted distributions are compared to data, a binned Poisson likelihood is used for all five SK data samples. The e-like samples use a 2D distribution in the reconstructed neutrino energy, $E_{\textrm{rec}}$, and the reconstructed neutrino angle with respect to the average beam direction, $\theta$. The $\mu$-like samples use a 1D distribution in the reconstructed neutrino energy.

For the rate+shape analysis, the likelihood for a pseudo-experiment is defined as the product of the likelihoods of the  $\bar{\nu}$ mode single-ring e-like sample, $\lambda_{\bar{\nu}_{e}}$, and the control samples, $\lambda_{\textrm{c}}$. The test statistic is then calculated as in equation (3), by averaging this likelihood over samples of the systematic parameter space, $a_i$. When the generated distribution of the test statistic is calculated, $\lambda_{\bar{\nu}_{e}}$ is compared to the pseudo-experiment data, $E$, and $\lambda_{\textrm{c}}$ is compared to data, $D$; when the test statistic for the real data is calculated, both likelihoods are compared to data.

\begin{equation}
\chi^2 \left( \beta \right) = -2\ln \left[ \frac{1}{N} \sum_{i=1}^{N} \right.
    \lambda_{\bar{\nu}_{e}} \left( \beta, \textbf{a}_{i}; E \right) \cdot 
    \lambda_{\textrm{c}} \left( \beta, \textbf{a}_{i}; \textrm{D} \right) 
\left. \vphantom{\sum_{i=1}^{N}} \right]
\label{eq:chisq_marg}
\end{equation}


An independent, complementary analysis uses the kinematic variable of outgoing lepton momentum, $p_{\textrm{l}}$ instead of reconstructed neutrino energy, and additionally uses weighting of pseudo-experiments instead of rejection sampling. Both analyses were found to give consistent test statistic distributions and therefore $p$-values. 

\begin{figure}[!ht]
	\includegraphics[width=\linewidth]{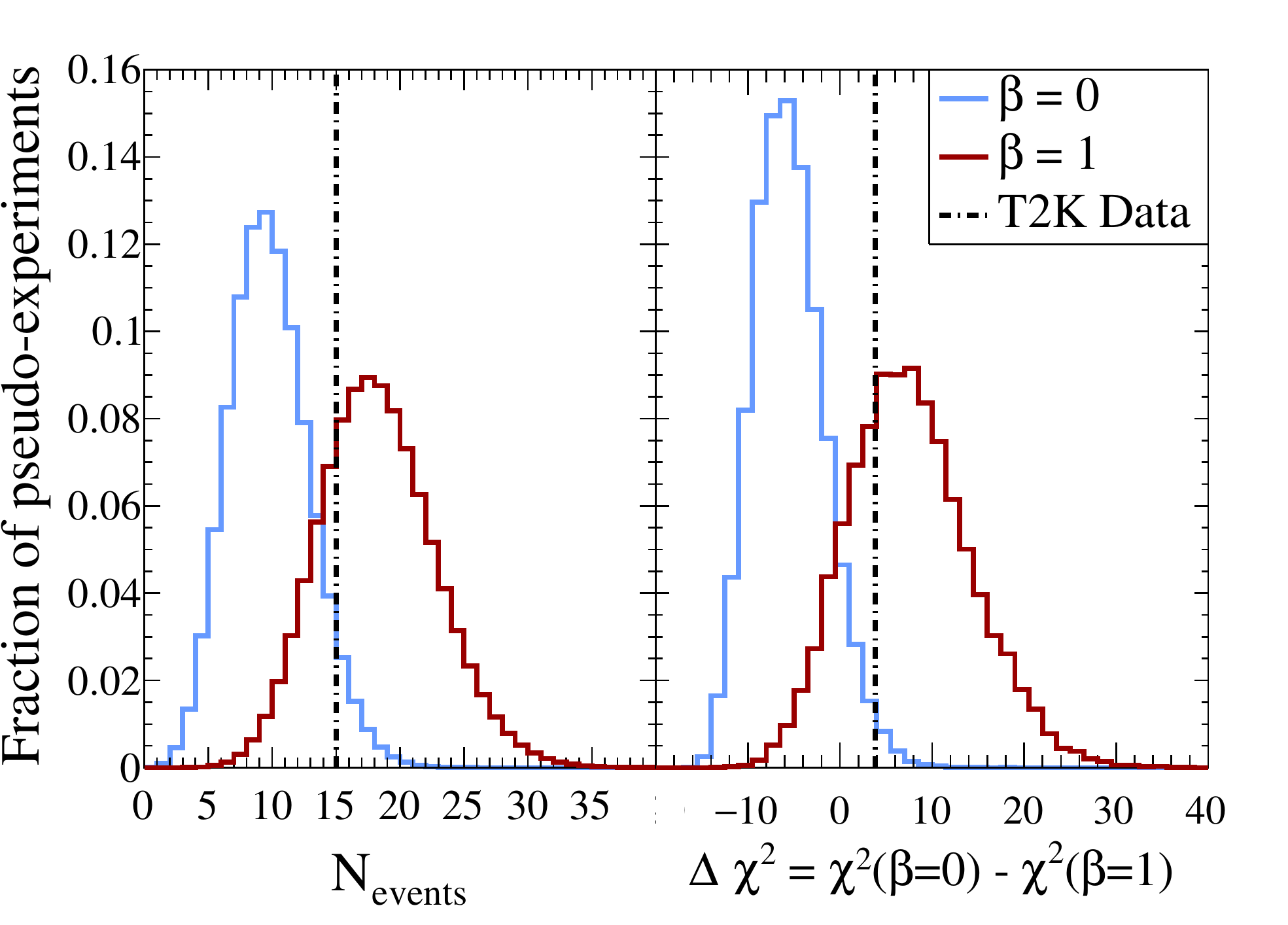}
	\hspace*{-0.5cm}
	\caption{Test statistic distributions taken from the $\beta=0$ and $\beta=1$ pseudo-experiment ensembles for the rate-only analysis (left) and rate+shape analysis (right). Here $\text{N}_{\text{events}}$ denotes the number of observed events in the $\nub$ mode single-ring e-like sample.}
	\label{fig:nuebar_test_stat_dists}
\end{figure}

The distributions of the rate-only and rate+shape test statistics for the $\beta=0$ and $\beta=1$ hypotheses are shown in Fig. \ref{fig:nuebar_test_stat_dists}. These distributions are integrated from the data test statistic to obtain right(left)-tailed $p$-values for the $\beta = 0(1)$ hypothesis. The observed number of events in the $\nub$ mode single-ring e-like sample in SK was \NskObservedNueBarRunOneNineD, compared to a prediction of \NskAsimovNueBarRunOneNineD. The observed data $\Delta \chi^2$ value in the rate+shape analysis was \DeltachisqDataRunOneNineD\ and the prediction was \DeltachisqAsimovRunOneNineD. The resulting $p$-values are shown in Tab. \ref{tab:data_pvals_R1-9d}. Both the rate-only and rate+shape analyses disfavor the no-$\nueb$-appearance hypothesis ($\beta=0$) more than the PMNS $\nueb$ appearance hypothesis ($\beta=1$). Compared to the prediction, a slightly weaker exclusion of the no $\nueb$ appearance hypothesis ($\beta=0$) is observed due to observing fewer events than expected. The rate+shape analysis gives a stronger observed exclusion of both hypotheses than the rate-only analysis, due to the extra shape information used to discredit each hypothesis.



\begin{table}
	\centering
	\caption{$p$-values and significance of the $\beta = 0$ and $\beta = 1$ hypotheses using both the rate-only and rate+shape analyses}
	\begin{tabular}{|c|c|c|c|c|c|}
		\hline
		\multirow{ 2}{*}{$\beta$} & \multirow{ 2}{*}{Analysis}	& \multicolumn{ 2}{|c|}{$p$-value} & \multicolumn{ 2}{|c|}{Significance ($\sigma$)} \\
		\cline{3-6}
		                          &                             & Expected & Observed            & Expected & Observed \\
		\hline
		\multirow{ 2}{*}{0}  & rate-only   & \pvalRateOnlyBetaZeroAsimovARunOneNineD  & \pvalRateOnlyBetaZeroDataRunOneNineD  & \sigmaRateOnlyBetaZeroAsimovARunOneNineD  & \sigmaRateOnlyBetaZeroDataRunOneNineD    \\
		\cline{2-6}
			   				 & rate+shape  & \pvalRateShapeBetaZeroAsimovARunOneNineD  & \pvalRateShapeBetaZeroDataRunOneNineD & \sigmaRateShapeBetaZeroAsimovARunOneNineD  & \sigmaRateShapeBetaZeroDataRunOneNineD   \\
		\hline
		\multirow{ 2}{*}{1}  & rate-only   & \pvalRateOnlyBetaOneAsimovARunOneNineD  & \pvalRateOnlyBetaOneDataRunOneNineD   & \sigmaRateOnlyBetaOneAsimovARunOneNineD  & \sigmaRateOnlyBetaOneDataRunOneNineD     \\
		\cline{2-6}
			   				 & rate+shape  & \pvalRateShapeBetaOneAsimovARunOneNineD  & \pvalRateShapeBetaOneDataRunOneNineD  & \sigmaRateShapeBetaOneAsimovARunOneNineD  & \sigmaRateShapeBetaOneDataRunOneNineD    \\
		\hline
	\end{tabular}
	\label{tab:data_pvals_R1-9d}
\end{table}

\begin{figure}[htbp]
  \centering
  \includegraphics[ width=\linewidth]{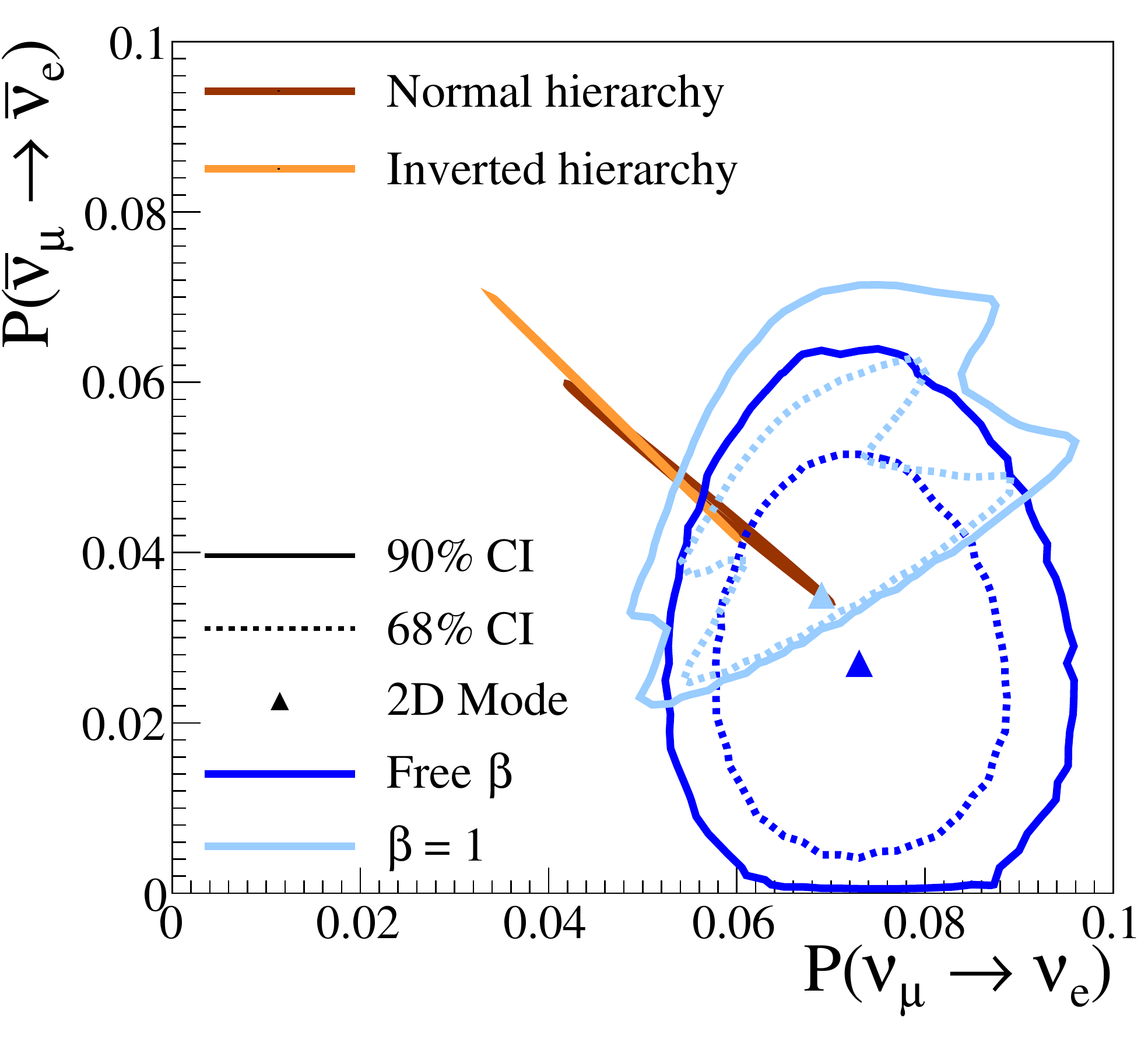}
  \caption{Bi-probability credible interval comparison between the standard fit
  constrained by the PMNS (light blue) model and the non-PMNS fit with the free
  $\beta$ parameterization (dark blue). The maximum posterior density point is
  marked as the 2D mode. The narrow T2K prediction ovals for inverse and normal
  mass hierarchies are in brown and orange respectively. In the ellipses, the bottom right corresponds to $\delta_{CP} = -\pi/2$, top left to $\delta_{CP} = \pi/2$, and the middle to $\delta_{CP} = 0, \pm \pi$. All probabilities are calculated at 600~MeV.}
  \label{fig:free-beta:resultcomparison}
\end{figure}

{\it Continuous $\beta$}\textemdash A complementary analysis allows $\beta$ to be a free parameter, which allows for a continuum of non-PMNS models, rather than only the single $\beta=0$ no-\nueb-appearance case. The impact of this analysis is shown in the parameter space of \pnue vs \pnueb. 
Varying \dcp at a fixed energy creates an ellipse with a negatively sloping major axis in the biprobability phase space. Switching the mass hierarchy shifts the center of the ellipse along the \pnue=~$-$\pnueb axis. The other oscillation parameters shift the ellipses along the identity line in the biprobability space. Two ellipses are shown in Fig.~\ref{fig:free-beta:resultcomparison} in orange and brown, with the input oscillation parameter values taken from the $\beta = 1$ fit; the eccentricity of the ellipses is very large for the T2K experiment, which makes them appear like lines. In the ellipses, the bottom right corresponds to $\delta_{CP} = -\pi/2$, top left to $\delta_{CP} = \pi/2$, and the middle to $\delta_{CP} = 0, \pm \pi$.

Credible interval contours (68\% and 90\%) are produced by a Bayesian Markov Chain Monte Carlo (MCMC) for the standard, fixed $\beta = 1$ parameterization and the new non-PMNS continuous-$\beta$ parameterization. These are shown in Fig~\ref{fig:free-beta:resultcomparison}. Both the credible intervals and the expectation ellipses are calculated with a neutrino beam energy fixed to 600~MeV.

In the fit with $\beta$ fixed to 1, two lobes appear in the contours, which correspond to the two mass hierarchies: the upper lobe to the inverted hierarchy, and the lower to the normal hierarchy. These lobes coincide with the maximally $CP$-violating \dcp value regions of the two T2K expectation ovals, shown in brown (inverse hierarchy) and orange (normal hierarchy). The width of the credible intervals comes mainly from the uncertainties in \stot and \stt, and height from \dcp and the mass hierarchy.

The free $\beta$ fit explores a larger area, especially in \pnueb, which is expected; the lower number of \nueb than \nue candidate events leads to a higher uncertainty in \pnueb, when not constrained by the PMNS model; additionally, the two probabilities are now decoupled due to the additional $\beta$ parameter, giving an independent result for both probabilities. These credible intervals can be used to compare other neutrino oscillation models against the fit constrained by the PMNS model and against the free $\beta$ fit that represents the information given by the T2K data with additional freedom. 

The 90\% and the 68\% credible intervals from both continuous-$\beta$ and PMNS-constrained fits significantly overlap. There is good agreement between the two fits, showing consistency between T2K data and the PMNS model. Additionally, the value of $\beta$ is consistent with 1 (90\% credible interval [0.3,1.06]), when marginalizing over all other oscillation parameters.

{\it Conclusions}\textemdash The T2K collaboration has searched for $\bar\nu_e$ appearance in a $\bar\nu_\mu$ beam using a data set twice as large as in its previous searches. The data have been analyzed within two frameworks, and have been compared to predictions with either no $\nueb$ appearance or $\nueb$ appearance as expected from the PMNS model prediction. In both frameworks, the data are consistent with the presence of $\bar\nu_e$ appearance and no significant deviation from the PMNS prediction is seen. Using full rate and shape information, the no-appearance scenario is disfavored with a significance of 2.40 standard deviations. 

{\it Acknowledgements}\textemdash We thank the J-PARC staff for superb accelerator performance. We thank the CERN NA61/SHINE Collaboration for providing valuable particle production data. We acknowledge the support of MEXT, Japan; NSERC (Grant No. SAPPJ-2014-00031), NRC and CFI, Canada; CEA and CNRS/IN2P3, France; DFG, Germany; INFN, Italy; National Science Centre (NCN) and Ministry of Science and Higher Education, Poland; RSF (Grant \#19-12-00325) and Ministry of Science and Higher Education, Russia; MINECO and ERDF funds, Spain; SNSF and SERI, Switzerland; STFC, UK; and DOE, USA. We also thank CERN for the UA1/NOMAD magnet, DESY for the HERA-B magnet mover system, NII for SINET4, the WestGrid, SciNet and CalculQuebec consortia in Compute Canada, and GridPP in the United Kingdom. In addition, participation of individual researchers and institutions has been further supported by funds from ERC (FP7), "la Caixa” Foundation (ID 100010434, fellowship code LCF/BQ/IN17/11620050), the European Union’s Horizon 2020 Research and Innovation programme under the Marie Sklodowska-Curie grant agreement no. 713673 and H2020 Grant No. RISE-GA644294-JENNIFER 2020; JSPS, Japan; Royal Society, UK; and the DOE Early Career program, USA.

\bibliographystyle{apsrev4-2}
\bibliography{apssamp}

\providecommand{\noopsort}[1]{}\providecommand{\singleletter}[1]{#1}%
\begin{thebibliography}{28}%
\makeatletter
\providecommand \@ifxundefined [1]{%
 \@ifx{#1\undefined}
}%
\providecommand \@ifnum [1]{%
 \ifnum #1\expandafter \@firstoftwo
 \else \expandafter \@secondoftwo
 \fi
}%
\providecommand \@ifx [1]{%
 \ifx #1\expandafter \@firstoftwo
 \else \expandafter \@secondoftwo
 \fi
}%
\providecommand \natexlab [1]{#1}%
\providecommand \enquote  [1]{``#1''}%
\providecommand \bibnamefont  [1]{#1}%
\providecommand \bibfnamefont [1]{#1}%
\providecommand \citenamefont [1]{#1}%
\providecommand \href@noop [0]{\@secondoftwo}%
\providecommand \href [0]{\begingroup \@sanitize@url \@href}%
\providecommand \@href[1]{\@@startlink{#1}\@@href}%
\providecommand \@@href[1]{\endgroup#1\@@endlink}%
\providecommand \@sanitize@url [0]{\catcode `\\12\catcode `\$12\catcode
  `\&12\catcode `\#12\catcode `\^12\catcode `\_12\catcode `\%12\relax}%
\providecommand \@@startlink[1]{}%
\providecommand \@@endlink[0]{}%
\providecommand \url  [0]{\begingroup\@sanitize@url \@url }%
\providecommand \@url [1]{\endgroup\@href {#1}{\urlprefix }}%
\providecommand \urlprefix  [0]{URL }%
\providecommand \Eprint [0]{\href }%
\providecommand \doibase [0]{https://doi.org/}%
\providecommand \selectlanguage [0]{\@gobble}%
\providecommand \bibinfo  [0]{\@secondoftwo}%
\providecommand \bibfield  [0]{\@secondoftwo}%
\providecommand \translation [1]{[#1]}%
\providecommand \BibitemOpen [0]{}%
\providecommand \bibitemStop [0]{}%
\providecommand \bibitemNoStop [0]{.\EOS\space}%
\providecommand \EOS [0]{\spacefactor3000\relax}%
\providecommand \BibitemShut  [1]{\csname bibitem#1\endcsname}%
\let\auto@bib@innerbib\@empty
\bibitem [{\citenamefont {Fukuda}\ \emph {et~al.}(1998)\citenamefont {Fukuda}
  \emph {et~al.}}]{PhysRevLett.81.1562}%
  \BibitemOpen
  \bibfield  {author} {\bibinfo {author} {\bibfnamefont {Y.}~\bibnamefont
  {Fukuda}} \emph {et~al.} (\bibinfo {collaboration} {Super-Kamiokande
  Collaboration}),\ }\href {https://doi.org/10.1103/PhysRevLett.81.1562}
  {\bibfield  {journal} {\bibinfo  {journal} {Phys. Rev. Lett.}\ }\textbf
  {\bibinfo {volume} {81}},\ \bibinfo {pages} {1562} (\bibinfo {year}
  {1998})}\BibitemShut {NoStop}%
\bibitem [{\citenamefont {Ahmad}\ \emph {et~al.}(2001)\citenamefont {Ahmad}
  \emph {et~al.}}]{PhysRevLett.87.071301}%
  \BibitemOpen
  \bibfield  {author} {\bibinfo {author} {\bibfnamefont {Q.~R.}\ \bibnamefont
  {Ahmad}} \emph {et~al.} (\bibinfo {collaboration} {SNO Collaboration}),\
  }\href {https://doi.org/10.1103/PhysRevLett.87.071301} {\bibfield  {journal}
  {\bibinfo  {journal} {Phys. Rev. Lett.}\ }\textbf {\bibinfo {volume} {87}},\
  \bibinfo {pages} {071301} (\bibinfo {year} {2001})}\BibitemShut {NoStop}%
\bibitem [{\citenamefont {Abe}\ \emph {et~al.}(2011{\natexlab{a}})\citenamefont
  {Abe} \emph {et~al.}}]{PhysRevLett.107.041801}%
  \BibitemOpen
  \bibfield  {author} {\bibinfo {author} {\bibfnamefont {K.}~\bibnamefont
  {Abe}} \emph {et~al.} (\bibinfo {collaboration} {T2K Collaboration}),\ }\href
  {https://doi.org/10.1103/PhysRevLett.107.041801} {\bibfield  {journal}
  {\bibinfo  {journal} {Phys. Rev. Lett.}\ }\textbf {\bibinfo {volume} {107}},\
  \bibinfo {pages} {041801} (\bibinfo {year} {2011}{\natexlab{a}})}\BibitemShut
  {NoStop}%
\bibitem [{\citenamefont {An}\ \emph {et~al.}(2012)\citenamefont {An} \emph
  {et~al.}}]{PhysRevLett.108.171803}%
  \BibitemOpen
  \bibfield  {author} {\bibinfo {author} {\bibfnamefont {F.~P.}\ \bibnamefont
  {An}} \emph {et~al.},\ }\href
  {https://doi.org/10.1103/PhysRevLett.108.171803} {\bibfield  {journal}
  {\bibinfo  {journal} {Phys. Rev. Lett.}\ }\textbf {\bibinfo {volume} {108}},\
  \bibinfo {pages} {171803} (\bibinfo {year} {2012})}\BibitemShut {NoStop}%
\bibitem [{\citenamefont {Maki}\ \emph {et~al.}(1962)\citenamefont {Maki},
  \citenamefont {Nakagawa},\ and\ \citenamefont {Sakata}}]{maki1962remarks}%
  \BibitemOpen
  \bibfield  {author} {\bibinfo {author} {\bibfnamefont {Z.}~\bibnamefont
  {Maki}}, \bibinfo {author} {\bibfnamefont {M.}~\bibnamefont {Nakagawa}}, and\
  \bibinfo {author} {\bibfnamefont {S.}~\bibnamefont {Sakata}},\ }\href@noop {}
  {\bibfield  {journal} {\bibinfo  {journal} {Progress of Theoretical Physics}\
  }\textbf {\bibinfo {volume} {28}},\ \bibinfo {pages} {870} (\bibinfo {year}
  {1962})}\BibitemShut {NoStop}%
\bibitem [{\citenamefont {Pontecorvo}(1968)}]{pontecorvo1968neutrino}%
  \BibitemOpen
  \bibfield  {author} {\bibinfo {author} {\bibfnamefont {B.}~\bibnamefont
  {Pontecorvo}},\ }\href@noop {} {\bibfield  {journal} {\bibinfo  {journal}
  {Sov. Phys. JETP}\ }\textbf {\bibinfo {volume} {26}},\ \bibinfo {pages} {165}
  (\bibinfo {year} {1968})}\BibitemShut {NoStop}%
\bibitem [{\citenamefont {Abe}\ \emph {et~al.}(2014)\citenamefont {Abe} \emph
  {et~al.}}]{abe2014observation}%
  \BibitemOpen
  \bibfield  {author} {\bibinfo {author} {\bibfnamefont {K.}~\bibnamefont
  {Abe}} \emph {et~al.},\ }\href@noop {} {\bibfield  {journal} {\bibinfo
  {journal} {Phys. Rev. Lett.}\ }\textbf {\bibinfo {volume} {112}},\ \bibinfo
  {pages} {061802} (\bibinfo {year} {2014})}\BibitemShut {NoStop}%
\bibitem [{\citenamefont {Abe}\ \emph {et~al.}(2017)\citenamefont {Abe} \emph
  {et~al.}}]{PhysRevD.96.092006}%
  \BibitemOpen
  \bibfield  {author} {\bibinfo {author} {\bibfnamefont {K.}~\bibnamefont
  {Abe}} \emph {et~al.} (\bibinfo {collaboration} {The T2K Collaboration}),\
  }\href {https://doi.org/10.1103/PhysRevD.96.092006} {\bibfield  {journal}
  {\bibinfo  {journal} {Phys. Rev. D}\ }\textbf {\bibinfo {volume} {96}},\
  \bibinfo {pages} {092006} (\bibinfo {year} {2017})}\BibitemShut {NoStop}%
\bibitem [{\citenamefont {Acero}\ \emph {et~al.}(2019)\citenamefont {Acero}
  \emph {et~al.}}]{acero2019first}%
  \BibitemOpen
  \bibfield  {author} {\bibinfo {author} {\bibfnamefont {M.}~\bibnamefont
  {Acero}} \emph {et~al.},\ }\href@noop {} {\bibfield  {journal} {\bibinfo
  {journal} {arXiv preprint arXiv:1906.04907}\ } (\bibinfo {year}
  {2019})}\BibitemShut {NoStop}%
\bibitem [{\citenamefont {Abe}\ \emph {et~al.}(2011{\natexlab{b}})\citenamefont
  {Abe} \emph {et~al.}}]{abe2011t2k}%
  \BibitemOpen
  \bibfield  {author} {\bibinfo {author} {\bibfnamefont {K.}~\bibnamefont
  {Abe}} \emph {et~al.} (\bibinfo {collaboration} {T2K Collaboration}),\
  }\href@noop {} {\bibfield  {journal} {\bibinfo  {journal} {Nucl. Instrum.
  Methods Phys. Res., Sect. A}\ }\textbf {\bibinfo {volume} {659}},\ \bibinfo
  {pages} {106} (\bibinfo {year} {2011}{\natexlab{b}})}\BibitemShut {NoStop}%
\bibitem [{\citenamefont {Abe}\ \emph {et~al.}(2018{\natexlab{a}})\citenamefont
  {Abe} \emph {et~al.}}]{PhysRevLett.121.171802}%
  \BibitemOpen
  \bibfield  {author} {\bibinfo {author} {\bibfnamefont {K.}~\bibnamefont
  {Abe}} \emph {et~al.} (\bibinfo {collaboration} {T2K Collaboration}),\ }\href
  {https://doi.org/10.1103/PhysRevLett.121.171802} {\bibfield  {journal}
  {\bibinfo  {journal} {Phys. Rev. Lett.}\ }\textbf {\bibinfo {volume} {121}},\
  \bibinfo {pages} {171802} (\bibinfo {year} {2018}{\natexlab{a}})}\BibitemShut
  {NoStop}%
\bibitem [{\citenamefont {Ferrari}\ \emph {et~al.}(2005)\citenamefont
  {Ferrari}, \citenamefont {Sala}, \citenamefont {Fasso},\ and\ \citenamefont
  {Ranft}}]{Ferrari:2005zk}%
  \BibitemOpen
  \bibfield  {author} {\bibinfo {author} {\bibfnamefont {A.}~\bibnamefont
  {Ferrari}}, \bibinfo {author} {\bibfnamefont {P.~R.}\ \bibnamefont {Sala}},
  \bibinfo {author} {\bibfnamefont {A.}~\bibnamefont {Fasso}}, and\ \bibinfo
  {author} {\bibfnamefont {J.}~\bibnamefont {Ranft}},\ }\href@noop {}
  {\bibfield  {journal} {\bibinfo  {journal} {CERN-2005-010}\ } (\bibinfo
  {year} {2005})}\BibitemShut {NoStop}%
\bibitem [{\citenamefont {Abgrall}\ \emph {et~al.}(2011)\citenamefont {Abgrall}
  \emph {et~al.}}]{Abgrall:2011ae}%
  \BibitemOpen
  \bibfield  {author} {\bibinfo {author} {\bibfnamefont {N.}~\bibnamefont
  {Abgrall}} \emph {et~al.} (\bibinfo {collaboration} {NA61/SHINE}),\ }\href
  {https://doi.org/10.1103/PhysRevC.84.034604} {\bibfield  {journal} {\bibinfo
  {journal} {Phys. Rev.}\ }\textbf {\bibinfo {volume} {C84}},\ \bibinfo {pages}
  {034604} (\bibinfo {year} {2011})}\BibitemShut {NoStop}%
\bibitem [{\citenamefont {Abgrall}\ \emph {et~al.}(2012)\citenamefont {Abgrall}
  \emph {et~al.}}]{PhysRevC.85.035210}%
  \BibitemOpen
  \bibfield  {author} {\bibinfo {author} {\bibfnamefont {N.}~\bibnamefont
  {Abgrall}} \emph {et~al.} (\bibinfo {collaboration} {NA61/SHINE
  Collaboration}),\ }\href {https://doi.org/10.1103/PhysRevC.85.035210}
  {\bibfield  {journal} {\bibinfo  {journal} {Phys. Rev. C}\ }\textbf {\bibinfo
  {volume} {85}},\ \bibinfo {pages} {035210} (\bibinfo {year}
  {2012})}\BibitemShut {NoStop}%
\bibitem [{\citenamefont {Abgrall}\ \emph {et~al.}(2016)\citenamefont {Abgrall}
  \emph {et~al.}}]{Abgrall:2016jif}%
  \BibitemOpen
  \bibfield  {author} {\bibinfo {author} {\bibfnamefont {N.}~\bibnamefont
  {Abgrall}} \emph {et~al.} (\bibinfo {collaboration} {NA61/SHINE}),\ }\href
  {https://doi.org/10.1140/epjc/s10052-016-4440-y} {\bibfield  {journal}
  {\bibinfo  {journal} {Eur. Phys. J.}\ }\textbf {\bibinfo {volume} {C76}},\
  \bibinfo {pages} {617} (\bibinfo {year} {2016})}\BibitemShut {NoStop}%
\bibitem [{\citenamefont {Abe}\ \emph {et~al.}(2013)\citenamefont {Abe} \emph
  {et~al.}}]{Abe:2012av}%
  \BibitemOpen
  \bibfield  {author} {\bibinfo {author} {\bibfnamefont {K.}~\bibnamefont
  {Abe}} \emph {et~al.} (\bibinfo {collaboration} {T2K}),\ }\href
  {https://doi.org/10.1103/PhysRevD.87.012001, 10.1103/PhysRevD.87.019902}
  {\bibfield  {journal} {\bibinfo  {journal} {Phys. Rev.}\ }\textbf {\bibinfo
  {volume} {D87}},\ \bibinfo {pages} {012001} (\bibinfo {year} {2013})},\
  \bibinfo {note} {[Addendum: Phys. Rev.D87,no.1,019902(2013)]}\BibitemShut
  {NoStop}%
\bibitem [{\citenamefont
  {Posiada\l{}a-Zezula}(2017)}]{Posiadala-Zezula:2017ivt}%
  \BibitemOpen
  \bibfield  {author} {\bibinfo {author} {\bibfnamefont {M.}~\bibnamefont
  {Posiada\l{}a-Zezula}},\ }\href
  {https://doi.org/10.1088/1742-6596/888/1/012064} {\bibfield  {journal}
  {\bibinfo  {journal} {J. Phys. Conf. Ser.}\ }\textbf {\bibinfo {volume}
  {888}},\ \bibinfo {pages} {012064} (\bibinfo {year} {2017})}\BibitemShut
  {NoStop}%
\bibitem [{\citenamefont {Zambelli}\ \emph {et~al.}(2017)\citenamefont
  {Zambelli}, \citenamefont {Fiorentini}, \citenamefont {Vladisavljevic} \emph
  {et~al.}}]{zambelli2017towards}%
  \BibitemOpen
  \bibfield  {author} {\bibinfo {author} {\bibfnamefont {L.}~\bibnamefont
  {Zambelli}}, \bibinfo {author} {\bibfnamefont {A.}~\bibnamefont
  {Fiorentini}}, \bibinfo {author} {\bibfnamefont {T.}~\bibnamefont
  {Vladisavljevic}},  \emph {et~al.},\ }\href@noop {} {\bibfield  {journal}
  {\bibinfo  {journal} {J. Phys. Conf. Ser.}\ }\textbf {\bibinfo {volume}
  {888}},\ \bibinfo {pages} {012067} (\bibinfo {year} {2017})}\BibitemShut
  {NoStop}%
\bibitem [{\citenamefont {Hayato}(2009)}]{hayato2009neutrino}%
  \BibitemOpen
  \bibfield  {author} {\bibinfo {author} {\bibfnamefont {Y.}~\bibnamefont
  {Hayato}},\ }\href@noop {} {\bibfield  {journal} {\bibinfo  {journal} {Acta
  Phys. Polon.}\ }\textbf {\bibinfo {volume} {40}},\ \bibinfo {pages} {2477}
  (\bibinfo {year} {2009})}\BibitemShut {NoStop}%
\bibitem [{\citenamefont {Nieves}\ \emph {et~al.}(2004)\citenamefont {Nieves},
  \citenamefont {Amaro},\ and\ \citenamefont {Valverde}}]{PhysRevC.70.055503}%
  \BibitemOpen
  \bibfield  {author} {\bibinfo {author} {\bibfnamefont {J.}~\bibnamefont
  {Nieves}}, \bibinfo {author} {\bibfnamefont {J.~E.}\ \bibnamefont {Amaro}},
  and\ \bibinfo {author} {\bibfnamefont {M.}~\bibnamefont {Valverde}},\ }\href
  {https://doi.org/10.1103/PhysRevC.70.055503} {\bibfield  {journal} {\bibinfo
  {journal} {Phys. Rev. C}\ }\textbf {\bibinfo {volume} {70}},\ \bibinfo
  {pages} {055503} (\bibinfo {year} {2004})}\BibitemShut {NoStop}%
\bibitem [{\citenamefont {Gran}\ \emph {et~al.}(2013)\citenamefont {Gran} \emph
  {et~al.}}]{PhysRevD.88.113007}%
  \BibitemOpen
  \bibfield  {author} {\bibinfo {author} {\bibfnamefont {R.}~\bibnamefont
  {Gran}} \emph {et~al.},\ }\href {https://doi.org/10.1103/PhysRevD.88.113007}
  {\bibfield  {journal} {\bibinfo  {journal} {Phys. Rev. D}\ }\textbf {\bibinfo
  {volume} {88}},\ \bibinfo {pages} {113007} (\bibinfo {year}
  {2013})}\BibitemShut {NoStop}%
\bibitem [{\citenamefont {Nieves}\ \emph {et~al.}(2011)\citenamefont {Nieves},
  \citenamefont {Simo},\ and\ \citenamefont {Vacas}}]{PhysRevC.83.045501}%
  \BibitemOpen
  \bibfield  {author} {\bibinfo {author} {\bibfnamefont {J.}~\bibnamefont
  {Nieves}}, \bibinfo {author} {\bibfnamefont {I.~R.}\ \bibnamefont {Simo}},
  and\ \bibinfo {author} {\bibfnamefont {M.~J.~V.}\ \bibnamefont {Vacas}},\
  }\href {https://doi.org/10.1103/PhysRevC.83.045501} {\bibfield  {journal}
  {\bibinfo  {journal} {Phys. Rev. C}\ }\textbf {\bibinfo {volume} {83}},\
  \bibinfo {pages} {045501} (\bibinfo {year} {2011})}\BibitemShut {NoStop}%
\bibitem [{\citenamefont {Valverde}\ \emph {et~al.}(2006)\citenamefont
  {Valverde}, \citenamefont {Amaro},\ and\ \citenamefont
  {Nieves}}]{VALVERDE2006325}%
  \BibitemOpen
  \bibfield  {author} {\bibinfo {author} {\bibfnamefont {M.}~\bibnamefont
  {Valverde}}, \bibinfo {author} {\bibfnamefont {J.}~\bibnamefont {Amaro}},
  and\ \bibinfo {author} {\bibfnamefont {J.}~\bibnamefont {Nieves}},\ }\href
  {https://doi.org/https://doi.org/10.1016/j.physletb.2006.05.053} {\bibfield
  {journal} {\bibinfo  {journal} {Physics Letters B}\ }\textbf {\bibinfo
  {volume} {638}},\ \bibinfo {pages} {325 } (\bibinfo {year}
  {2006})}\BibitemShut {NoStop}%
\bibitem [{\citenamefont {Gran}(2017)}]{gran2017model}%
  \BibitemOpen
  \bibfield  {author} {\bibinfo {author} {\bibfnamefont {R.}~\bibnamefont
  {Gran}},\ }\href@noop {} {\bibfield  {journal} {\bibinfo  {journal} {arXiv
  preprint arXiv:1705.02932}\ } (\bibinfo {year} {2017})}\BibitemShut {NoStop}%
\bibitem [{\citenamefont {{Day~}}\ and\ \citenamefont
  {McFarland}(2012)}]{PhysRevD.86.053003}%
  \BibitemOpen
  \bibfield  {author} {\bibinfo {author} {\bibfnamefont {M.}~\bibnamefont
  {{Day~}}}and\ \bibinfo {author} {\bibfnamefont {K.}~\bibnamefont
  {McFarland}},\ }\href {https://doi.org/10.1103/PhysRevD.86.053003} {\bibfield
   {journal} {\bibinfo  {journal} {Phys. Rev. D}\ }\textbf {\bibinfo {volume}
  {86}},\ \bibinfo {pages} {053003} (\bibinfo {year} {2012})}\BibitemShut
  {NoStop}%
\bibitem [{\citenamefont {Kabirnezhad}(2018)}]{PhysRevD.97.013002}%
  \BibitemOpen
  \bibfield  {author} {\bibinfo {author} {\bibfnamefont {M.}~\bibnamefont
  {Kabirnezhad}},\ }\href {https://doi.org/10.1103/PhysRevD.97.013002}
  {\bibfield  {journal} {\bibinfo  {journal} {Phys. Rev. D}\ }\textbf {\bibinfo
  {volume} {97}},\ \bibinfo {pages} {013002} (\bibinfo {year}
  {2018})}\BibitemShut {NoStop}%
\bibitem [{\citenamefont {Abe}\ \emph {et~al.}(2018{\natexlab{b}})\citenamefont
  {Abe} \emph {et~al.}}]{Abe:2018wpn}%
  \BibitemOpen
  \bibfield  {author} {\bibinfo {author} {\bibfnamefont {K.}~\bibnamefont
  {Abe}} \emph {et~al.} (\bibinfo {collaboration} {T2K}),\ }\href
  {https://doi.org/10.1103/PhysRevLett.121.171802} {\bibfield  {journal}
  {\bibinfo  {journal} {Phys. Rev. Lett.}\ }\textbf {\bibinfo {volume} {121}},\
  \bibinfo {pages} {171802} (\bibinfo {year} {2018}{\natexlab{b}})},\ \Eprint
  {https://arxiv.org/abs/1807.07891} {arXiv:1807.07891 [hep-ex]} \BibitemShut
  {NoStop}%
\bibitem [{\citenamefont {Tanabashi}\ \emph {et~al.}(2018)\citenamefont
  {Tanabashi} \emph {et~al.}}]{PDG2018}%
  \BibitemOpen
  \bibfield  {author} {\bibinfo {author} {\bibfnamefont {M.}~\bibnamefont
  {Tanabashi}} \emph {et~al.} (\bibinfo {collaboration} {Particle Data
  Group}),\ }\href {https://doi.org/10.1103/PhysRevD.98.030001} {\bibfield
  {journal} {\bibinfo  {journal} {Phys. Rev. D}\ }\textbf {\bibinfo {volume}
  {98}},\ \bibinfo {pages} {030001} (\bibinfo {year} {2018})}\BibitemShut
  {NoStop}%
\end{thebibliography}%

\end{document}